\documentclass[useAMS]{mn2e}
\usepackage{graphicx}
\usepackage{amsmath}
\usepackage{amssymb}
\usepackage{color}
\usepackage{url}

\newcommand  \acc     {\ifmmode {\rm km\,s}^{-2} \else km\,s$^{-2}$\fi}

\newcommand  \ergs     {\ifmmode {\rm ergs\,s}^{-1} \else ergs s$^{-1}$\fi}
\newcommand  \ergcms   {\ifmmode {\rm erg~cm}^{-2}\,{\rm s}^{-1}
                        \else erg~cm$^{-2}$\,s$^{-1}$\fi}
\newcommand  \ergcmsA  {\ifmmode{\rm erg\,cm}^{-2}\,{\rm s}^{-1}\,{\rm\AA}^{-1}
                        \else ergs\,cm$^{-2}$\,s$^{-1}$\,\AA$^{-1}$\fi}
\newcommand  \ergcmsHz {\ifmmode{\rm ergs\,cm}^{-2}\,{\rm s}^{-1}\,{\rm Hz}^{-1}
                        \else ergs\,cm$^{-2}$\,s$^{-1}$\,Hz$^{-1}$\fi}
\newcommand  \phcms    {\ifmmode {\rm ph\,cm}^{-2}\,{\rm s}^{-1}
                        \else ph\,cm$^{-2}$\,s$^{-1}$\fi}
\newcommand  \phcmsA   {\ifmmode {\rm ph\,cm}^{-2}\,{\rm s}^{-1}\,{\rm\AA}^{-1}
                        \else ph\,cm$^{-2}$\,s$^{-1}$\,\AA$^{-1}$\fi}

\newcommand\aj{{AJ}}%
\newcommand\araa{{ARA\&A}}%
\newcommand\apj{{ApJ}}%
\newcommand\apjl{{ApJ}}%
\newcommand\apjs{{ApJS}}%
\newcommand\aap{{A\&A}}%
\newcommand\aapr{{A\&A~Rev.}}%
\newcommand\aaps{{A\&AS}}%
\newcommand\mnras{{MNRAS}}%
\newcommand\pasp{{PASP}}%
\title[Bursts from Nearby Flaring Stars]
{Fast radio bursts may originate from nearby flaring stars}
\author[A. Loeb, Y. Shvartzvald, D. Maoz]
{Abraham Loeb$^{1,2}$, Yossi Shvartzvald$^{2}$, Dan Maoz$^{2}$\\
$^{1}$Institute for Theory and Computation, 
Harvard University, Cambridge, MA 03210, USA\\
$^{2}$School of Physics and Astronomy, Tel-Aviv University, Tel-Aviv 69978,
Israel} \date{\today}

\begin{document}

\maketitle

\label{firstpage}

\begin{abstract}
Six cases of fast radio bursts (FRBs) have recently been
 discovered. The FRBs are bright ($\sim
0.1-1~{\rm Jy}$) and brief ($\sim 1~{\rm ms}$) 
pulses of radio emission with dispersion measures (DMs) 
that exceed Galactic values, and hence FRBs have been interpreted to
be at cosmological distances. We propose, instead, that FRBs are rare
eruptions of flaring main-sequence stars within $\sim 1$~kpc. Rather
than associating their excess DM with the intergalactic medium, we
relate it to a blanket of coronal plasma around their host star.  We
have monitored at optical bands the stars within the radio beams of
three of the known FRBs. In one field, we find a bright (V=13.6~mag)
variable star (0.2~mag peak-to-trough) with a main-sequence G-type
spectrum and a period $P=7.8$~hr, likely a W-UMa-type contact
binary. 
Analysis of our data outside of the FRB beams indicates a
$5\%$ probability of finding, at random, a variable star of this
brightness and amplitude within the FRB beams, but this could still be
a chance coincidence.  We find no unusual variable stars in the other
two FRB fields. Further observations are needed to investigate if
similar nearby ($\lesssim 800$~pc) stars are the sources of FRBs.
 \end{abstract}




\newpage

\section{Introduction}

Over the past 6 years, six cases of fast radio bursts
(FRBs) have been discovered (Lorimer et al. 2007; Keane et al. 2012;
Thornton et al. 2013; see Lorimer et al. 2013 for an overview). 
The origin of
this potentially new population of sources 
is particularly enigmatic given that the FRBs are
bright ($\sim 0.1-1~{\rm Jy}$) and brief ($\sim 1~{\rm ms}$)
pulses of $\sim 1~{\rm GHz}$ radio emission, with a
total estimated rate of $\sim 10^4~{\rm day^{-1}}$ over the whole sky
(Thornton et al. 2013). The FRBs were each detected only once over the
$\sim 5$~min duration of radio coverage of each field. 
No counterparts have been detected at other
wavelengths, but the analysis of the radio data and the 
discovery of the FRBs occurred years after the events. 
The FRBs exhibit unusually high dispersion measures
(DMs). The first FRB discovered by Lorimer et al. (2007) showed
DM$=375~{\rm cm^{-3}~pc}$; the second identified by Keane et
al. (2012) had DM$=746~{\rm cm^{-3}~pc}$, and the remaining four FRBs
were found by Thornton et al. (2013) with DM$=550$--$1100~{\rm
cm^{-3}~pc}$. These DMs exceed the values expected from the column
density of interstellar electrons towards known radio sources, such as
pulsars, within the Milky Way, and hence FRBs have been inferred to
originate from extragalactic sources at cosmological distances. At
source redshifts $\gtrsim 0.1$ (or comoving distances of $\gtrsim 0.5~{\rm
Gpc}$), the intergalactic medium, combined with the DM of the sources'
host galaxies, could account for the excessive DM
values. 
In addition to the excess DM, two of the FRBs (one in Lorimer et
al. 2007, one in Thornton et al. 2013) show frequency-dependent pulse 
smearing, indicative of scattering by electrons along the
line of sight, at a level much larger than expected by a galactic
interstellar medium. The intergalactic medium has been invoked to
explain this scattering as well (e.g. Macquart \& Koay 2013).   

\section{Theoretical considerations for FRB emission}

A cosmological distance to FRBs makes their theoretical
interpretation challenging. Let us consider synchrotron emission from an
expanding shell, as commonly done for cosmological radio sources.  If
the emitting shell, of radius $R$,
 moves towards the observer with a Lorentz factor
$\Gamma$, then the pulse duration will be
$\Delta t\sim R/2c\Gamma^2$. For the millisecond durations of FRBs,
 this limits $R$ to a value
\begin{equation}
R\lesssim 6\times 10^7\Gamma^2 ~{\rm cm}\ \ .
\end{equation}
For synchrotron emission by a population of relativistic electrons
with a characteristic Lorentz factor $\gamma$ in the shell's
rest-frame, the observed frequency $\nu_{\rm
syn}\sim \Gamma \gamma^2 (eB/2\pi m_ec)\sim 1~{\rm GHz}$, implies (Rybicki \& Lightman 1979),
\begin{equation}
\Gamma\gamma^2(B/{\rm G}) \sim 3\times 10^2 \ \ , 
\end{equation}
where $B$ is the magnetic field strength in the shell's rest frame.
Using this result, the brightness of FRBs, $f_\nu\sim
N_e \Gamma [(e^3B/m_e c^2)/(4\pi d^2)]\sim 1~{\rm Jy}$, implies that, at a cosmological
distance $d\sim 1~{\rm Gpc}$, the (isotropically-equivalent) number of
radiating electrons $N_e$ is
\begin{equation}
N_e\sim 2\times 10^{52}\gamma^2 .
\end{equation}
The resulting optical depth for synchrotron self-absorption is then
(Waxman \& Loeb 1999),
\begin{equation}
\tau_{\rm syn}\approx \frac{1}{8\pi(\nu_{\rm
      syn}/\Gamma)^2}\left(\frac{N_e} 
{4\pi R^2 \gamma m_e}\right) \left(\frac{e^3B}{m_ec^2}\right)\gtrsim 10^{24}\gamma^{-1}\Gamma^{-3}.
\end{equation}
The emitting shell is thus highly opaque to the low-frequency
radiation it produces, which is able to escape only from its skin.
These considerations imply that it is difficult to explain the large
brightnesses and short timescales of FRBs at cosmological distances
with $\gamma,\Gamma\ll 10^6$ and with beamed synchrotron models that
are similar to those used to describe other extragalactic radio
sources, such as $\gamma$--ray burst afterglows (Meszaros 2013; Piran
2004) and active galactic nuclei (Begelman et al. 1980). Exotic
processes involving extremely relativistic outflows or coherent radio
emission by bunches of electrons (Katz 2013) near compact objects,
such as magnetars (Popov \& Postnov 2013), neutron star binaries
(Totani 2013), binary white-dwarf mergers (Kashiyama, Ioka, \&
Meszaros (2013), or supramassive neutron stars (Falcke \& Rezzolla
2013; Zhang 2013), must instead be considered.

However, even in these scenarios, regardless of the emission
mechanism, it is difficult to see how synchrotron absorption could be
avoided, if not in the emitting plasma itself, then in its immediate
environment. At cosmological distances, the required source luminosity $L$
exceeds the Eddington luminosity $L_E$  of a stellar progenitor of mass $M\sim 1$--$10M_\odot$ by orders of
magnitude, $L> ({\rm GHz\times Jy \times 4\pi \times Gpc^2}) \sim 10^{42}~{\rm ergs~s^{-1}}\gg L_{\rm E}=1.4\times 10^{38}(M/M_\odot)~{\rm erg~s^{-1}}$.
Thus, the repulsive radiation force dominates over gravity
and is likely to produce an electron-rich outflow outside the emission region. The
strong synchrotron absorption of the emitted GHz radiation by the likely
surrounding magnetized outflow has not been considered in the above-mentioned
models. Even for a modest electron column density and magnetic field strength in the outflow ,
one typically gets a synchrotron absorption optical depth (Rybicki \& Lightman 1979) that is much larger than unity 
at the low observed frequency of $\sim 1~{\rm GHz}$.

\section{FRBs from nearby stars?}

Given that the required number of radiating electrons increases with
source distance, it is much easier to satisfy the brightness
requirement of FRBs by invoking nearby Galactic sources. Since FRBs
were found by surveying regions of the sky with Galactic latitudes
$|b|=30^{\circ}-70^{\circ}$, 
a Galactic disk
source population must be within a distance of several 
stellar-disk scale heights, 
i.e. $\lesssim 1$ kpc from the Sun (about six orders
of magnitude closer than conjectured by the FRB discoverers). A halo
population may, of course, be somewhat more distant. The
local stellar mass density is $0.085\pm0.010~ M_\odot$~pc$^{-3}$ 
(Mcmillan 2011),  
which gives $\sim 10^8$ stars within this disk volume, most of them with
mass $\sim
0.5 M_\odot$, i.e. K- and M-type dwarfs. 

Some flaring dwarf stars are already known to produce coherent radio
bursts with rise times shorter than $5$ milliseconds and $\sim 1$~GHz
flux densities of a fraction of a Jy (Lang et al. 1983; Lang \& Wilson
1986; G{\"u}del et al. 1989, Bastian et al.  1990). These bursts are
thought to be produced by the cyclotron maser mechanism, in which
coherent radio emission with $f_\nu\propto N_e^2$ occurs due to
bunching of the emitting electrons as they gyrate around the magnetic
field of the host star (G{\"u}del 2002; Treumann 2006; Matthews 2013).
If FRB eruptions are produced at the bottom of the coronae of their
host stars, or cause coronal mass ejections of their own (Drake et
al. 2013), the radiation they produce is likely to pass through a
plasma blanket with a characteristic electron column density $\gtrsim
10^{10}~{\rm cm^{-3}}\times R_\odot\sim 300~{\rm cm^{-3}~pc}$, as
needed to explain the observed DMs.

The same plasma blanket could also produce the pulse smearing observed
in some FRBs. For the characteristic electron density at the base of
the corona of a flaring star, $n_e\sim 10^{10}~{\rm cm}^{-3}$, the
plasma frequency is $\nu_p\sim 0.9$~GHz, below but comparable to the observed
photon frequency in FRBs. The turbulence generated in this plasma
during a flare could therefore produce scattering of the radio waves
at large angles, and hence the observed, $\sim 1$~ms, temporal
smearing through the associated time delays between different light
paths. Since, in this scenario, the scattering screen is positioned on
top of the source, the distant thin screen approximation used for
smearing by an intergalactic screen (Goodman \& Narayan 1989; Macquart
\& Koay 2013) is not applicable. 
However, if the source of an FRB flare is 
a localized reconnection event,  then the point source approximation may still apply.
We expect a low frequency cutoff in the emission spectrum of FRBs. Its
detection would allow measuring 
the plasma frequency, and hence density,
of  the stellar coronae in which FRBs are produced.

Stellar magnetic flaring activity is often seen in two circumstances
-- in young, low-mass, stars, and in solar-mass contact binaries. In
either case, the source stars can be identified by means of their
optical variability, whether flaring, periodic, or quasi-periodic.
Walkowicz et al. (2011) have used data from the {\it Kepler} mission
to characterise the statistics of flaring activity in low-mass
stars. They find that the large majority of flaring stars are M-type
stars. Furthermore, flaring activity is confined to M stars only
within a height of $z\lesssim 100$~pc above the Galactic plane, and at
$z=40$~pc almost 50\% of M stars flare. This height dependence arises
because only young M stars are magnetically active. Once such stars
age, they are dynamically scattered to the larger scale height of the
stellar disk.  Thus, if FRB sources are flaring M stars, they will be
within $\sim 100$~pc. The active M stars from {\it Kepler} generally
also show quasi-periodic ``quiescent variability'', due to coverage of
their surfaces by spots, with peak-to-peak amplitudes of order a few
percent (Basri et al. 2011). Thus, such stars may be identifiable by
means of their optical variability.

The second class of common stars with flaring activity are
binaries of the W Ursa Majoris (W UMa) type, consisting of two stars, usually
F to K-type, in contact within a common envelope. Such systems
show periodic optical photometric variability, with $P\sim 8$~hr periods, due
to the combination of mutual eclipses by the components, and their
ellipsoidal tidal distortion (e.g., Rucinski 1997; 
Molnar, van Noord, \& Steenwyk 2013). 
 
As a test of our local FRB-source hypothesis, we
have therefore monitored in the optical the fields of
three of the six known FRBs, in search of unusual optical variables.

\section{Observations and analysis of FRB fields}

We observed the fields of three of the known FRBs using the Wise
Observatory 1-m telescope in Mitspe Ramon, Israel. The targets, dates
(UT at beginning of night), and mode of observation, are listed in
Table~\ref{table:obsjournal}. For imaging, we used LAIWO, an
8k$\times$8k CCD mosaic that covers a non-contiguous $1~{\rm deg}^2$
field of view at the Cassegrain focus of the telescope. Johnson $V$
and $I$ filters were used on some nights, and only the $V$ band on
others.  Exposure times varied from night to night, from 1 to 10~min
per exposure. On UT August 2, 2013, observations of Landolt (1992)
photometric standard fields were interspersed during the night with
the FRB fields. A photometric solution for these data using the {\tt
meastan} routine written by one of us (D.M.) in IRAF\footnote{Image
Reduction and Analysis Facility (http://iraf.noao.edu/)} was used to
calibrate photometrically the stars in each field to 0.03~mag
accuracy.

\begin{table}
\center
\caption{Journal of Observations}
\begin{tabular}{l|c|c}
\hline
\hline
FRB Field &  RA(2000)DEC  & UT Dates (2013, mmdd)\\
\hline
110220 &  22:34:00 $-$12:24:00 & 0728-0730, 0802, 0902, 0906, 0908\\
110703 &  23:30:00 $-$02:52:00 & 0728-0730, 0802, 0815$^*$,   \\
       & 		      & 0826, 0901-0909, 0910$^*$\\
120127 &  23:15:00 $-$18:25:00 & 0728-0730, 0802, 0815$^*$, 0826, 0901, \\
       & 		      & 0903, 0905, 0907, 0909, 0910$^*$ \\
\hline
\end{tabular}
\noindent $^{*}$ Spectroscopic observations; otherwise imaging.
\label{table:obsjournal}
\end{table}

As further detailed below, photometric analysis of the imaging data revealed
variability of one of the bright stars 
 in the FRB 110703 field
(Thornton et al. 2013).
Spectra of this star
were obtained at Wise with the FOSC, an all-transmission
low-resolution spectrograph.
A $10''$ slit was used to minimize slit losses due to guiding
and atmospheric refraction (given the southern declinations, 
the star had to be observed at rather large air masses, of 1.2-1.8).
Individual exposures were 15-20~min, and  spectra covered the
3800--7550~\AA~ range with a spectral resolution full-width at half
maximum of ${\rm FWHM}\sim 10$~\AA.
Standard imaging and spectroscopic data reductions were performed using IRAF.

Aperture photometry was performed using SExractor (Bertin \& Arnouts
1996) on all of the stars and at all observed epochs in the LAIWO
fields toward each FRB. We used differential photometry to obtain a
light curve for every source, and from the multiple epochs we measured
the mean and root-mean-square ({\it rms}) magnitude. Figure~1 plots
the {\it rms} versus the mean for the stars in the LAIWO field that
includes FRB 110703, obtained on the nights UT June 28-30, 2013. Stars
within the $14'$ diameter FRB error circle are plotted in red, and the
others in black. The locus of the measurements shows that, for bright
stars approaching the CCD saturation limit at $V\sim 12$~mag, our
photometry is stable to a few millimag {\it rms} precision. Precision
deteriorates with fading flux with the dependence expected from
Poisson statistics, reaching a signal-to-noise ratio S/N$\approx 5$
for $V=19.5$~mag stars.

\begin{figure}
\label{fig:rmsmag}
\center
\begin{tabular}{c}
\includegraphics[width=0.5\textwidth]{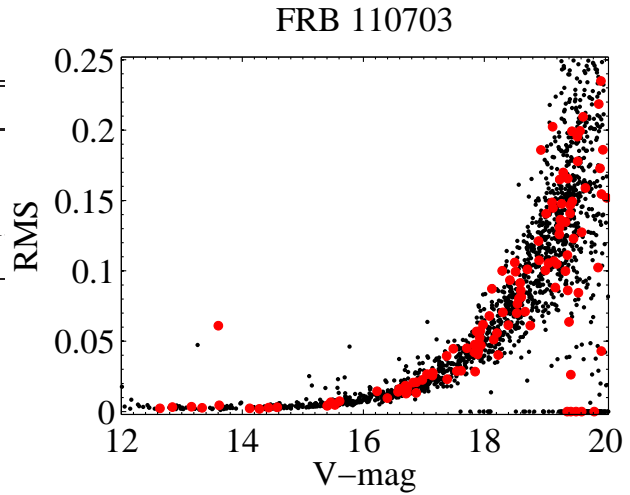}
\end{tabular}
\caption{Root mean square versus  mean $V$-band magnitudes for the stars in the $1{\rm deg}^2$ LAIWO field that includes FRB 110703.
Stars within the $14'$-diameter FRB FWHM
 are plotted with 
large filled red circles, and other stars as  small black dots.
A bright variable star within the FRB beam, V233003-024818, 
is conspicuous at $V=13.6$~mag.
}
\end{figure}

Data for all candidate variable stars lying above the
locus were inspected for artifacts such as cosmic-ray events,
CCD edge effects, diffraction spikes from bright stars, and confusion
due to crowding effects. All real variables were further analysed using 
 difference image analysis with the
PYSIS (Albrow et al. 2009) program, to confirm the reality of the variations, and to
refine the accuracy of the light curves. We describe below our
findings in each field.

\subsection{FRB 110703}

Figure~1 identifies a variable bright ($V=13.6$~mag) star in the FRB
110703 field, V233003-024818. Inspection of the light curve from these
nights, and subsequent photometry, shows clear periodic behaviour.
Figure~2 shows the star's location, well within the FRB error
circle. Figure~3 shows the $V$, $I$, and $V-I$ light curves, after
folding them with a period of $7.837$~hr. Fitting spectra from the
Pickles (1985) stellar library to the spectra that we have obtained,
we find the star's spectrum matches well a main-sequence,
solar-metallicity, early G-type star.  The star is detected in the
ultraviolet in {\it GALEX} data, with 1542~\AA~ and 2274~\AA~
magnitudes of $m_{\rm FUV}=23.77\pm0.49$ and $m_{\rm
NUV}=18.56\pm0.01$, respectively, and in 2MASS near-infrared data with
$J=12.42\pm0.02$~mag and $K=12.02\pm0.03$, all of which are at the
levels expected for a G star (Ducati et al. 2001; Bianchi et
al. 2007). It is undetected in the
{\it ROSAT} X-ray database.

\begin{figure}
\label{fig:finder1}
\center
\begin{tabular}{c}
\includegraphics[width=0.4\textwidth]{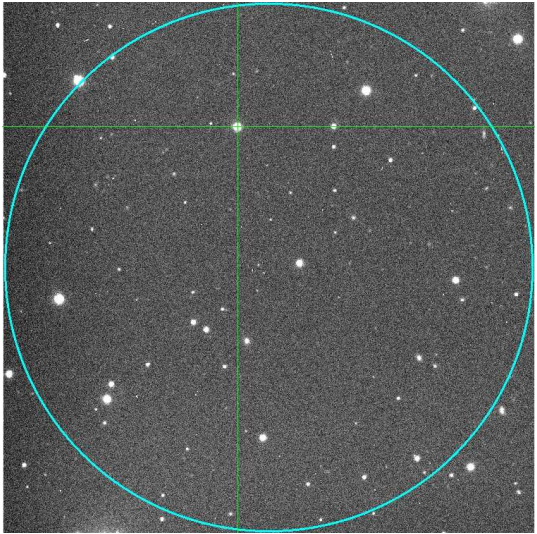}
\end{tabular}
\caption{FRB 110703 field. The circle shows the $14'$ diameter
  FRB beam's FWHM.
The periodic bright variable star, V233003-024818, is marked with a cross.
}
\end{figure}

\begin{figure}
\label{fig:lc1}
\center
\begin{tabular}{c}
\includegraphics[width=0.5\textwidth]{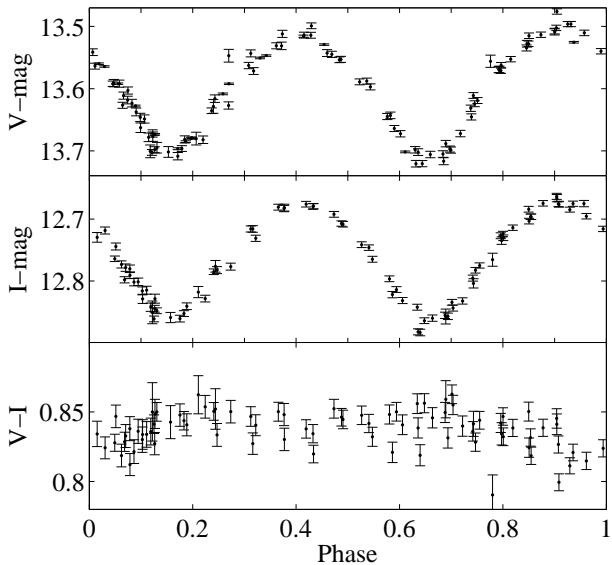}
\end{tabular}
\caption{Folded light curve for the variable star, V233003-024818, in
  the field of FRB 110703,
with a period of $7.837$~hr. 
}
\end{figure}

The period, amplitude, light-curve shape, low level of colour
variability, and spectrum, are all suggestive of a typical W UMa
system, consisting of two G-type stars in a contact binary. From the
0.22~mag amplitude and the similar depths of the two minima, the
system's orbital pole has a low inclination to our line of sight (as
opposed to edge-on systems, where mutual eclipses of the two
components result in amplitudes of factor $\sim 2$, and differing
depths of the two minima). Scaling the {\it ROSAT} X-ray count rates
of about $1~{\rm s}^{-1}$, seen in nearby W UMa systems (McGale, Pye,
\& Hodgekin 1996), by the optical magnitude differences, a count rate
of only $\sim 0.003$~s$^{-1}$ X-ray photons would be expected in
V233003-024818, consistent with the present non-detection.

The system may, alternatively, be a single, ellipsoidally distorted, G
star, in orbit with an unseen lower-mass companion -- an M star or a
compact stellar remnant. We exclude the possibility that the periodic
variability results from a single pulsating star with half of the
above period (i.e. $P=3.9$~hr) because: {\it (i)} main-sequence G
stars do not pulsate with such periods and amplitudes; {\it (ii)}
pulsating stars change colour over a period, as their radii and
temperatures oscillate.
Finally, it is possible, but seems unlikely, that this is a single G
star rotating near breakup speed, with a pattern of spots that is
stable over at least several months, and that the spots somehow
produce the smooth lightcurve observed.

We conclude that V233003-024818 is very likely a W UMa
system. Assuming that, at maximum, we are seeing the luminosities of
two main-sequence G2 stars, the distance to the system is about
800~pc. For its Galactic latitude of $b=59^{\circ}$, this corresponds
to a height $z=600$~pc above the Galactic plane, 1.5 times the disk
scale height of W UMa systems (Rucinski 1997).

\subsection{FRB 110220} 
In this FRB field, we have found no conspicuous variables. At
magnitudes fainter than $V\sim 16$, variables are seen above the {\it
rms} noise locus in the {\it rms} versus mean diagram, in this, as in
the other fields. However, they occur as frequently within the FRB
beams as outside them. While one of them might be a flare star that
was the source of the FRB, our current data cannot identify it as
such. This null-detection field hosted the brightest
among the Thornton et al. (2013) FRBs, with a 1.3 Jy flux density at
peak.

We note that the FRB beam for this event includes also a 8.8~mag star,
HD 213735, which is saturated in our data, and hence we could not test
it for variability. We are currently monitoring this star with a
smaller telescope and short exposures.

\subsection{FRB 120127}

As in the FRB 110220 field, 
in this field as well we find no unusual variable stars, other than 
the common fainter examples.  

\section{Discussion}

We have argued that FRB sources need not be exotic events at
cosmological distances, but rather could be due to extreme magnetic
activity in nearby Galactic stars.  To test this hypothesis, we have
monitored the stars in the directions of three FRBs, in search of
unusual optically variable stars that could signal stellar FRB
sources. In one field, we have found a likely W UMa contact binary
system at a distance of 800~pc. The total solid angle of sky that we
have monitored outside of the FRB beams is 20 times larger than the
solid angle inside the beams. In the non-FRB-beam region, we have
found only one other bright and periodic variable comparable to
V233003-024818 (also in the field of FRB 110703, but outside the radio
beam). None of the much fainter variables that we have found are
periodic. This suggests a $\sim 5\%$ probability for the chance
coincidence of V233003-024818 within one of the FRB beams. A similar
estimate of this probability comes from considering the frequency of W
UMa systems among solar-like stars.  Rucinski (1997) estimates a
Solar-neighborhood space density of W UMa systems of $1.5\times
10^{-4}~{\rm pc}^{-3}$. Assuming a G-star space density of $\sim
0.03~{\rm pc}^{-3}$, about one in 200 G-stars is in a W UMa
system. Within the three FRB beams that we have monitored, there are
12 stars brighter than 14 mag and with $0.6 <V-I< 1.2$, similar to
late F to early G stars. This again suggests a $\sim 5\%$ chance
probability for a W UMa system within one of the FRB beams.
 
Interestingly, W UMa systems likely have the right coronal properties
necessary to explain the DMs of the observed FRBs, in the range of
400-1100 cm$^{-3}$~pc.  For example, Gondoin (2004) has observed and
analysed the spectral evolution of one of the frequent X-ray flaring
events in VW Cephei, a well-studied W UMa system . The deduced
electron density, $n_e> 6.5\times 10^{10}$~cm$^{-3}$, combined with a
flare loop length of $3\times 10^{10}$~cm, comparable to the size of
the flaring primary star, gives an electron column density of $N_{\rm
col, e}\sim 2\times 10^{21}$~cm$^{-2}=700$~cm$^{-3}$~pc.  The X-ray
temperatures are sufficient to propel the gas beyond the escape speed
from the star, $v\sim 400$~km~s$^{-1}$.  The frequent ejection of
material with such column densities and scale lengths implies a mass
loss of order $1~M_\odot$ over the 1~Gyr estimated lifetime of the
contact phase (Eker, Demircan, \& Bilir 2008). This is consistent with
the mass losses inferred by Yildiz \& Dogan (2013) who compare
evolutionary models of W UMa binaries to known systems.
  
 We have found no unusually
variable stars in the two other FRB fields that we have observed.
Our results present several possibilities. First, the presence of the
periodic variable in one FRB beam area could be, after all, a chance
coincidence. There is an element of {\it a posteriori} statistics in
our analysis, since we set out on a ``fishing expedition'' for unusual
optical variables, without defining in advance exactly what types of
objects or properties would constitute such variables. The true FRB
sources could therefore be some other stars in the fields that we have
not noted (perhaps from among the fainter variable stars seen in the
fields), or the sources could be extragalactic after all, as
postulated in previous papers.  Alternatively, if the W UMa system
really is the FRB 110703 source, 
this would raise the question of why a W UMa
system is not seen in the other two FRB beams as well. Perhaps 
diverse stellar systems and processes can lead to an FRB, unless,
e.g., there are W UMa systems in all three FRB beams, but in two cases
they are undetected as such because their orbits are oriented nearly
pole-on to our line of sight.

If W UMa systems are indeed the sources of FRBs, their known space
density allows us to estimate the mean recurrence time of FRBs from a
given source. Within the 800~pc distance to V233003-024818, and assuming a
scale height of 450~pc for W UMa systems (Rucinski 1997), 
there are roughly $1.5\times 10^5$ such systems. The FRB rate of
$10^4$~day$^{-1}$ estimated by Thornton et al. (2013) then implies 
that an FRB will recur at the same position about once every two weeks, on
average. This prediction can be tested by continuous longterm
radio monitoring of FRB sites. This will soon be possible with
wide-field radio surveys such as the Murchison Widefield Array (Trott,
Tingay, \& Wayth 2013). 

The degree of isotropy of FRB events is yet unclear. The four Thornton
et al. (2013) FRBs were discovered by searching 24\% of a
high-galactic-latitude survey region, $30^\circ<|b|<70^\circ$.  Thornton
et al. (2013) also report that no FRBs were found in a mid-latitude,
$|b|<15^\circ$, region. Assuming an isotropic distribution, based on the
high-latitude numbers, $7^{+4}_{-3}$ FRBs were expected, under ideal
conditions, in the full mid-latitude survey.  Since only 23\%
of the mid-latitude data have been analyzed, 
the expectation value at mid-latitudes
is 1.6 FRBs. And since the ``Milky Way foreground would reduce this
rate, with increased sky temperature, scattering, and dispersion, for
surveys close to the Galactic plane'', the expectation value at
$|b|<15^\circ$ is likely $\lesssim 1$, and the lack of detections there,
at this stage, is not surprising.  If FRBs originate from a Galactic
disk population, more than one scale height away (where the
scale-height is 450~pc for W UMa systems or 100~pc for active young M
stars), then future FRB detections should gradually show a preference
for the Galactic plane, with a cumulative number above photon flux
$f_\nu$ that scales as $N(>f_\nu)\propto f_\nu^{-1}$, due to the
two-dimensional geometry of the disk convolved with the intrinsic flux
distribution per source. For closer Galactic sources, less than a
scale-height away, there should be little difference in isotropy
compared to a scenario involving cosmological distances.
 
\section*{Acknowledgments}
We thank S. Kaspi, T. Mazeh, and G. Nelemans for valuable discussions
and input. The referee, J.-P. Macquart, is thanked for useful
comments.  
A.L. acknowledges support from the Sackler Professorship
by Special Appointment at Tel Aviv University. This work was supported in part
by NSF grant AST-1312034 (for A.L.); and by a grant from the US-Israel
Binational Science Foundation and Grant 1829/12 of the I-CORE program
of the PBC and the Israel Science Foundation (for D.M.).


\end{document}